\documentclass[10pt]{article}
\usepackage{amsmath} \usepackage{amssymb}
\newcommand{\be}{\begin{equation}}
\newcommand{\ee}{\end{equation}}
\newcommand{\bea}{\begin{eqnarray}}
\newcommand{\eea}{\end{eqnarray}}
\newcommand{\bean}{\begin{eqnarray*}}
\newcommand{\eean}{\end{eqnarray*}}
\newcommand{\barr}{\begin{array}}
\newcommand{\earr}{\end{array}}

\newcommand{\nonum}{\nonumber}
\newcommand{\txi}{\tilde{\Xi}}

\newcommand{\tj}{\tilde{j}}

\newcommand{\trho}{\tilde{\rho}}

\newcommand{\wt}{\widehat T}

\date{\today}

\begin{document} 
\begin{titlepage}
\begin{flushright} UFIFT-HEP-04-12 \\
\end{flushright}
\vskip .5cm
\centerline{\LARGE{\bf {Continuous Spin Representations }}}
\vskip .2cm
\centerline{\LARGE{\bf  from Group Contraction }}
\vskip 1cm
\date{\today}
\centerline{\bf Abu M. Khan\footnote{E-mail: amas@phys.ufl.edu} and Pierre Ramond\footnote{E-mail:
ramond@phys.ufl.edu}} \vskip .5cm \centerline{\em  Institute for Fundamental Theory,}
\centerline{\em Department of Physics, University of Florida} \centerline{\em Gainesville FL
32611, USA} \vskip 1.5cm
\begin{abstract}
We consider how the continuous spin representation (CSR) of the Poincar\'e group in four
dimensions can be generated by dimensional reduction. The analysis uses the front-form little
group in five dimensions, which must yield the Euclidean group $E(2)$, the little group of the
CSR. We consider two cases, one is the single spin massless representation of the Poincar\'e group
in five dimensions, the other is the infinite component Majorana equation, which describes an
infinite tower of massive states in five dimensions. In the first case, the double singular limit
$j\,,\,R\to \infty$, with $j/R$ fixed, where $R$ is the Kaluza-Klein radius of the fifth
dimension, and $j$ is the spin of the particle in five dimensions, yields the CSR in four
dimensions. It amounts to the In\"on\"u-Wigner contraction, with the inverse K-K radius as
contraction parameter. In the second case, the CSR appears only by taking a triple singular limit,
where an internal coordinate of the Majorana theory goes to infinity, while leaving its ratio to
the KK radius fixed.
\end{abstract}
\vskip .5cm \noindent \vfill
\end{titlepage}

\noindent
\section{Introduction}
As shown in Wigner's classic work~\cite{WIGNER} in four dimensions, there are four types of
irreducible representations of the Poincar\'e group. Two describe massless and massive elementary
particles with definite helicity and spin, respectively. However, Nature does not seem to use the
other two:  one that describes particles with space-like momenta~\cite{BARGMANN}, tachyons which
move faster than the speed of light, and the others  that describe massless states with an
infinite number of integer or half-odd integer unit-spaced helicities, dubbed  by Wigner
Continuous Spin Representations(CSRs).

When tachyons occur in field and string theories, it is as symptoms of an unstable theory, a
malady whose cure is known:  by  shifting the vacuum through spontaneous symmetry breaking as in
field and string field theory, or by extending  to supersymmetry as in  string theory.

There is no analogous cure for the CSR, for which the obstacles are indeed formidable: negative
norm states, non-locality,  and acausality~\cite{hirata}, and according to Wigner
himself~\cite{WIGNER2}, infinite heat capacity of the vacuum.

The bosonic CSR necessarily  contains one massless graviton, but accompanied by an infinite tower
of massless helicity states. Therefore it should  be viewed in the context of theories that extend
General Relativity, such as  M- or String Theories, where a naive application of the infinite
slope limit to their spectrum  leads to an infinite array of massless states with unit-spaced
helicities.  Like the tachyonic representation, the CSR may also be symptomatic of a diseased
theory, but is there a cure for it?

Such considerations merit further studies. In a previous paper~\cite{lars}, Wigner's bosonic and
fermionic CSR in four dimensions were found to be supersymmetric partners of one another.
Interestingly, the supersymmetric CSR does not have infinite heat capacity as it cancels between
bosons and fermions. We also showed how to generalize them to higher dimensions by using Dirac's
front form (a.k.a. light-cone) where the CSRs are linked to representations of the Euclidian group
in the transverse plane in any dimensions.

The purpose of  this paper is to seek CSR's in familiar mathematical structures which contain the
Poincar\'e group in four dimensions: the Conformal group $SO(4,2)$, the Poincar\'e group in
higher-dimensions, and in the de Sitter and Anti-de Sitter groups in five dimensions. The latter
require group contraction, introduced by In\"on\"u and Wigner(IW)~\cite{iw}, in order to generate
the Euclidean group from the rotation group. We show that, while the CSR do not appear as
representations of the Conformal group~\cite{FRONSDAL, YAO}, they can be generated from the
five-dimensional Poincar\'e group by a combination of group contraction and Kaluza-Klein~\cite{KK}
dimensional reduction. We apply the same procedure to a more complicated theory, Majorana's
infinite component wave equation \cite{majorana} in five dimensions. We do not consider how string
theories in higher dimension can yield CSR's in lower dimensions, our considerations indicate a
connection between group contraction and their infinite slope limit.

\section{Front Forms of the Poincar\'e and Conformal Algebras}
In this section, we briefly review the light-cone forms of the Poincar\'e and Conformal
algebras in arbitrary dimensions.
\subsection{Poincar\'e Algebra}
In $d$ dimensions, the generators of the Poincar\'e algebra satisfy the commutation
relations,
\bea \nonumber [P^\mu , P^\nu] & = & 0 \ , \\ \nonumber [ M^{\mu \nu} , P^\sigma ] & = &
i (\eta^{\mu \sigma} P^\nu - \eta^{\nu \sigma} P^\mu)\ , \\ \nonumber [ M^{\mu \nu} ,
M^{\alpha \beta} ] & = & i (\eta^{\mu \alpha} M^{\nu \beta} + \eta^{\alpha \nu} M^{\beta
\mu} + \eta^{\nu \beta} M^{\mu \alpha} + \eta^{\beta \mu} M^{\alpha \nu}) \ , \eea
where $\eta_{\mu\nu}=(-1,1,\cdots,1)$ and $\mu , \nu ,\alpha , \beta = 0,\cdots , (d-1)$.

Introduce the light-cone coordinates~\cite{db,lars},
$$ x^\pm ~ = ~ \frac{1}{\sqrt{2}}(t \pm x^{d-1}) \quad {\rm and}
\quad p^\pm ~ = ~ \frac{1}{\sqrt{2}}(p^0 \pm p^{d-1}) \ , $$
where $i,j=1,2,\cdots , (d-2)$ are the transverse directions. The
commutation relations satisfied by the momenta and positions are
$$ [x^- , p^+ ] ~ = ~ -i \quad {\rm and} \quad [x^i , p^j ] ~ = ~
i \eta^{ij} \ . $$ Following Dirac, we set $x^+=0$ and use the mass-shell
condition to express its conjugate variable $p^-$ in terms of the remaining
variables. The translation generators become
$$ P^- ~ = ~ \frac{p^i p^i + M^2}{2p^+} \ , \quad P^+ ~ = ~ p^+ \ , \quad P^i ~ = ~ p^i \ ;$$
 $P^-$ is  called the light-cone Hamiltonian.
The Lorentz generators are given by
\bea  M^{+i} & = & -x^i p^+ \ , \label{m+i}\\
M^{+-} & = & - x^- p^+ \ , \label{m+-}\\ \label{mij} M^{ij} & = & x^i p^j - x^j p^i +
S^{ij} \ , \\ \label{m-i}  \quad M^{-i} & = &
x_{}^-p_{}^i-\frac{1}{2}\{x^i,P_{}^-\}+\frac{1}{p^+} (T_{}^i-p_{}^jS_{}^{ij})\ . \eea
The generators, $S^{ij}$, form the $SO(d-2)$ transverse light-cone little group and obey,
\bea \left[ S^{ij} , S^{kl} \right] & = & i \left( \eta^{ik} S^{jl} + \eta^{kj} S^{li} +
\eta^{jl} S^{ik} + \eta^{li} S^{kj} \right) \ . \eea
The $T^i$s transform as $SO(d-2)$ vectors, called the light-cone translation vectors
and satisfy
\bea
\left[ S^{ij} \ , \ T^k\right] & = & i\left( \eta^{ik} T^j - \eta^{jk} T^i \right) \
, \label{lc-vector}\\ \left[ T^i \ , \ T^j \right] & = & i M^2 S^{ij} \ .
\label{lc-commutator}\eea
The difference between the various representations is easily understood in
terms of these variables:
\begin{itemize}

\item
 When $M\ne 0$, we can divide Eq.(\ref{lc-commutator}) by $M$, so that

$$S^{ij}~~{\rm and}~~\frac{T^i}{M}$$
generate the massive little group $SO(d-1)$. The representations are therefore
labelled by the massive little group.

\item  When $M=0$, the $T^i$ are still $SO(d-2)$ vectors, but they commute with one another.
 This leads us to consider two cases:
\begin{enumerate}

\item $T^i=0$, yields the regular massless representation labelled by a representation of the
$SO(d-2)$ little group generated by $S^{ij}$, with a finite number of helicity states.

\item $T^i\ne0$. In this case, the $S^{ij}$ and $T^i$s form the inhomogeneous light-cone little
group in $(d-2)$ dimensions, which has infinite-dimensional representations, yielding the
 continuous spin massless representations of the Poincar\' e group.

\end{enumerate}
\end{itemize}
\subsection{Conformal Algebra}
The Poincar\' e group is a subgroup of the Conformal group $SO(d,2)$ which contains dilatation and
conformal transformations. The concept of mass makes sense in the conformal sense only if it is
zero, to preserve scale invariance. Thus one expects massless representations of the Poincar\' e
group to appear naturally in those of the conformal group.

It has been known~\cite{FRONSDAL, YAO} for some time that the CSR do not appear in this
decomposition. We can show it elegantly using light-cone coordinates, and  expressing the
generators of the Conformal algebra using the same variables as for the Poincar\' e algebra. We
need to add to the Poincar\'e generators the dilatation generator

\be D ~ = ~ \frac{1}{2}\left( x \cdot p + p\cdot x \right) \ ,  \ee
for the scale transformation and the special conformal transformations generated by,

\be K^\mu ~ = ~ 2 x_\nu M^{\mu \nu} + x^2 p^\mu  ~ = ~ 2 x^\mu D - x^2 p^\mu + 2 x^\nu S^{\mu \nu} \ ,
\ee
where $\mu =0,1, \cdots (d-1)$, and

\be M^{\mu \nu} ~ = ~ x^\mu p^\nu - x^\nu p^\mu + S^{\mu \nu} \ .  \ee In addition to those of the
Poincar\'e algebra, these two generators satisfy the following commutation relations,
$$ \left[ M^{\mu\nu} \ , \ D \right]  =  0 \ , \quad \left[ D \ , p^\mu \right]
= i p^\mu \ , \quad \left[ D \ , \ K^\mu \right] = -i K^\mu \ , $$ $$ \left[ M^{\mu \nu}
\ , \ K^\alpha \right] = i \left(\eta^{\mu \alpha} K^\nu - \eta^{\nu \alpha} K^\mu
\right) \ , \quad \left[ p^\mu \ , \ K^\nu \right] = -2i \left( \eta^{\mu \nu} D -
M^{\mu\nu} \right) \ . $$
To obtain the light-cone forms of the Dilatation and the Conformal translation generators, we first
express the dilation and conformal generators in light-cone coordinates,
\bea
D & = & \frac{1}{2}\left( - \{x^- ,p^+ \}  - \{ x^+ ,p^- \} + \{ x^i ,p^i \}
\right) \ ,\nonum \\   K^+ & = & -2 x^+ M^{+-} + x^2 p^+ + 2 x^i M^{+i} \ , \nonum \\   K^- &
= & 2 x^- M^{+-} + 2 x^i M^{-i} + x^2 p^- \ ,\nonum
\\ K^i & = & 2 x^- M^{+i} + 2 x^+ M^{-i} + 2 x^j M^{ij} + x^2 p^i  \ ,
\nonum \eea
where $i,j=1,\cdots , (d-2)$. Using the light-cone forms of the Poincar\'e generators
from Eqs.(\ref{m+i}-\ref{m-i}) and setting $x^+ =0$, $K^-$ and $K^i$
can be written as
\bea K^- & = & 2 x^- \left( -x^- p^+  + x^i p^i \right) - x^i x^i p^- + \frac{2}{p^+} \left( x^i
T^i - x^i p^j S^{ij} \right) \ , \\ K^i & = & 2 x^i \left( -x^- p^+  + x^i p^i \right) - x^j x^j
p^i + 2 x^j S^{ij} \ .  \eea
To check if these satisfy the algebra, we calculate the commutators $[P^- , K^i]$ and $[p^i ,
K^-]$. Substituting the light-cone forms, we find
\bea \label{p-ki} \left[ P^- \ , K^i \right] & = & 2i \left( x^- p^i - \frac{1}{2} \{ x^i \ , \
P^- \} - \frac{p^j}{p^+} S^{ij} \right) \ , \\ \label{pik-} -\left[ p^i \ , K^- \right] & = &  2i
\left( x^- p^i - \frac{1}{2} \{ x^i , P^- \} + \frac{1}{p^+} \left( T^i - p^j S^{ij} \right)
\right) \ . \eea
But to satisfy the Conformal algebra, these two commutators must be equal, requiring $T^i=0$.
Therefore, the front forms of the conformal generators are
\bea K^+ & = & - x^i x^i p^+ \ ,  \qquad K^i ~=~ 2 x^i D - x^j x^j p^i + 2 x^j
S^{ij} \ , \\  K^- &=& 2 x^- D - x^i x^i p^- - \frac{2}{p^+} \, x^i p^j S^{ij}
\ ,\\D &=& \frac{1}{2}\left( - \{x^- ,p^+ \} + \{ x^i ,p^i \} \right)
  \eea
This shows that  $T^i=0$, which imply regular massless representations, not the CSR.

We conclude that in order to generate a CSR by embedding the Poincar\'e algebra into larger
algebraic structures, we must consider singular limits. The CSRs require the transverse little
group to be the Euclidean group $E(2)$, the semi-direct product of translations and rotations. It
is the very same group that In\"on\"u and Wigner~\cite{iw} obtained by contracting the homogeneous
$SO(3)$ group. As this group is the massless transverse little group in five dimensions, we are
led to consider dimensional reduction as well as group contraction of the Poincar\'e group in
higher dimensions.


\section{Kaluza-Klein Reduction of the 5-d Poincar\'e Algebra}
Kaluza and  Klein~\cite{KK} reduced one dimension by putting it on to a circle of finite radius.
We apply it to the five-dimensional Poincar\'e algebra,  by putting the third direction on to a
circle of radius $R$,
$$ x^3 ~=~ x^3 + 2\pi R \ . $$
The generators act on functions of the form
 \be
\Phi(x) ~=~ \sum_{n} \Phi_n (x^\mu ) e^{inx^3/R}\ , \ee where $\mu = +,-,1,2$ and $n$ is the mode
number. The momentum along the third direction is  quantized

\be p^3 ~=~ \frac{n}{R} \ ,\label{P3} \ee
and the light-cone Hamiltonian becomes,
\be \label{massshell} P^- ~ = ~ \frac{p^i p^i + (p^3)^2+M^2}{2p^+} ~ \equiv ~ \frac{p^i p^i +
M^2_n}{2p^+} \ ,\ee where
\be M_n~=~M^2+\frac{n^2}{R^2}\ ,\ee
is the KK mass.  The single state of mass $M$ in five dimensions  generates an equally spaced
tower of states of mass $M_n$. To find their spin, observe that the two transverse boost
generators are
\be \label{boost} M^{-i} ~ = ~ x^- p^i - \frac{1}{2} \{x^i , P^- \} + \frac{1}{p^+} \left( {T}^i +
\frac{n}{R} S^{3i}- p^j S^{ij} \right) \ , \ee for $i=1,2$, leading us to define new light-cone
translation vectors
\be \widehat T^i ~ = ~ T^i + \frac{n}{R} S^{3i} \label{wideti} \ ,\ee which satisfy the algebra
\be [ S^{ij},\wt^k ]~=~ i(\eta^{ik}\wt^j - \eta^{jk}\wt^i )\ , \qquad [\wt^i,\wt^j]~=~ i M_n^2\,
S^{ij} \ .\label{widet-commutator} \ee
 The generators
$$ S^{12}\ , \qquad \frac{\wt^i}{M_n}\ , $$
form the $SO(3)$ light-cone little group in four dimensions. Analyzing the mass term,  the mass
$M_n$ increases with the mode number and gives the well-known infinite (KK) tower of masses
starting at  $M_0=M$. The remaining Lorentz generators of the four-dimensional group, $M^{+-},
M^{12}$ and $M^{+i}$ for $i=1,2$  are not changed.  Note that when evaluated at $x^3=2\pi R$, the
generators that rotate into the third direction become (for large $R$)  like the momenta

\bea \frac{M^{-3}}{R} & = & \frac{n}{R^2} x^- - 2 \pi p^- + \frac{1}{Rp^+} \left( T^3 - p^i
{S}^{3i} \right) \ , \\ \frac{M^{3i}}{R} & = & 2\pi p^i - \frac{n x^i}{R^2} + \frac{1}{R}S^{3i} \
, \\  \frac{M^{+3}}{R} & = & -2 \pi p^+ \ . \eea When the starting representation is massive
($M\ne 0$), the KK procedure yields massive representations in lower dimensions, but even if
$M=0$, the ground floor of a KK tower is the usual  massless representation in four dimensions,
not the CSR. In the simple limit $R\rightarrow\infty$, the KK mass tower collapses without any
sign of a CSR.
\section{In\"on\"u-Wigner Contraction }
This section presents the original In\"on\"u-Wigner~\cite{iw} contraction procedure. Starting from
the homogeneous $SO(3)$ group, they generated the Euclidean group $E(2)$, in a singular limit with
an arbitrary contraction parameter that tends to zero. They considered the algebra's generators,
identifying those which are well defined under contraction, as well as to the two-dimensional
wavefunction on which the Euclidean translation vectors act, which they show to be Bessel's
functions $J_n$.

They begin with the $SO(3)$ generators $L_i$ in the $(2l+1)$-dimensional representation, which
satisfy

$$[\,L^{}_i\,,\,L^{}_j\,]~=~i\,\epsilon_{ijk}\,L^{}_k\ ;\qquad  L^{}_i\,L^{}_i~=
~l\,(l+1)\ ,\qquad L^{}_3~=~m\ .$$
They introduce a contraction parameter $\epsilon$ and the vectors
\be T^{}_i~\equiv~ \epsilon\,L^{}_i\ . \ee They observe that in the limit $\epsilon\rightarrow 0$,
these, together with $L_3$ satisfy the commutation relations of the Euclidean algebra

\be [\,L^{}_3\,,\,T^{}_i\,]~=~\epsilon^{}_{ij}\,T^{}_j\ ,\qquad [\,T^{}_i\,,\,T^{}_j\,]~=~0\ . \ee
Multiplying the Casimir operator with $\epsilon^2$ yields

\be (\epsilon\,m)^2_{}+T^2_1+T^2_2~=~\epsilon^2_{}\,l\,(l+1)\ . \ee The In\"on\"u-Wigner
contraction is the double limit

\be l~\rightarrow~\infty\ ,\qquad \epsilon~\rightarrow~0\ ,\qquad \epsilon\,l ~
\equiv~\Xi~~\text{fixed}\ ,\ee in which the Euclidean translations have a finite length
\be T^2_1+T^2_2~=~\Xi^2_{}\ . \ee In their original paper, In\"on\"u and Wigner applied their
method to the Poincar\'e group, using the inverse of the speed of light as a contraction
parameter, and find it to contract to the non-relativistic Galilei group({\it but only if the
starting point is the tachyonic representation!}).

The representation function of $E(2)$ in a space labelled by the polar coordinates $\rho$ and
$\theta$ are of the form

\be \sum_n\,c^{}_n\,J^{}_n(\Xi\,\rho)\,e_{}^{in\,\theta}\ . \ee In the Appendix we show how to
obtain these results directly by imposing periodic boundary conditions \`a la Kaluza-Klein.

\section{The CSR  \`a la  Kaluza-Klein-In\"on\" u-Wigner}
The results of the preceding two sections hints on how to proceed: the contraction parameter
$\epsilon$ is simply the inverse of the Kaluza-Klein radius. We start from the usual massless
Poincar\'e group in five dimensions, and apply the KK procedure to the third direction. The
dynamical boosts are

\be M^{-i} ~ = ~ x_{}^- p_{}^i - \frac{1}{2}\, \{x_{}^i , P_{}^- \} + \frac{1}{p^+} \left(
\frac{n}{R}\, S_{}^{3i}- p_{}^j S_{}^{ij} \right)\ . \ee Comparison with the results of the
previous section leads us to identify the IW contraction parameter as

\be \epsilon~=~\frac{n}{R}\ . \ee
This yields the CSR boosts, where the translations are given by

\be \widehat T^i_{}~=~\frac{n}{R}\, S_{}^{3i}\ . \ee We mimic the IW procedure and consider the
Casimir operator of the $SO(3)$ light-cone little group, where $j$ is the spin of the massless
particle in five dimensions,

\be \nonumber (S^{12})^2 + (S^{23})^2 + (S^{31})^2 ~ = ~ j\,(j+1) \ . \ee Dividing both side by
$R^2$, we obtain

\be
 (\widehat T^1)^2 + (\widehat T^2)^2 ~ = ~\frac{n^2_{}\,j_{}^2}{R_{}^2} ~ \equiv ~ \Xi^2~\ne~ 0
 \ .\label{xi}
\ee The length of the translation vector is nothing but the Pauli-Lubanski Casimir operator of the
Poincar\'e group. This shows that it is possible to generate the CSR as long as one takes the
double limit

\be
R~\rightarrow \infty\ ,\qquad j~\rightarrow~\infty\ ,\qquad \frac{j}{R}~~\text{fixed}\ .
\ee
In the next section, we apply the same method to a more complicated theory.
%
%
\section{CSR from Majorana's Theory}
In this section, we consider the contraction of the five dimensional massive representations. A
simple, yet non-trivial model for the massive relativistic particle is the infinite component
Majorana theory, introduced by E. Majorana long ago\cite{majorana}. Its spectrum can be inferred
from Poincar\'e generators with a massive $SO(4)$ little group. Amusingly the mathematics of this
$SO(4)$ are exactly the same as those of Bohr's non-relativistic hydrogen atom. We want to apply
contraction to this theory and investigate under what circumstances a CSR can be generated.

In the following, we construct the mass operator and other generators of the little group in terms
of internal coordinates. This can be thought of the nonlinear realization of the homogeneous
$SO(4)$ group. We start with applying the contraction to the $SO(4)$ little group algebra and find
the Casimirs in five and four dimensions which labels the representations. We will then find the
conditions necessary for the existence of the CSRs. We will also find the representation of the
wavefunctions in four dimensions and compare with the IW paper.


\subsection{Majorana's Theory}
We start with the five dimensional light-cone realization of the dynamical Poincar\'e boosts

$$
M^{-\,i}_{}~=~x_{}^-\,p_{}^i-\,\frac{1}{2}\{\,x^i\,,\,P_{}^-\,\}+\frac{1}{p^+}\,
(T_{}^i-p_{}^jS_{}^{ij})\ ,$$ with $i\ , j=1,2,3$.

The little group generators are built in terms of three ``internal variables"  $\eta^i$ and their
conjugates $\pi^i$,

\be
[~\eta_{}^i~,~\pi_{}^j~]~=~i\,\delta_{}^{ij}\ ,\quad i,j=1,2,3\ ,\ee
in the form

\be
S_{}^{ij}~=~\epsilon_{}^{ijk}\,L_{}^k~=~\eta_{}^i\pi_{}^j-\eta_{}^j\pi_{}^i\ ,\ee
and with the non-linear Laplace-Runge-Lenz(LRL) vectors

\be T_{}^{i}~=~\epsilon_{}^{ijk}\,\pi_{}^j\,L_{}^k- \mu \,\frac{\eta_{}^i}{\eta}-i\,\pi_{}^i\ ,
\ee where $\eta=\sqrt{\eta^i\eta^i}$ and $\mu$ is a mass parameter (we have set the ``electron
charge" to one). They obey the commutation relations

\be
[\, T_{}^{i}\,,\,T_{}^{j}\,]~=~i(-2\mu \,H)\,\epsilon^{ijk}_{}\,L^k_{}\ ,\label{llc}
\ee
where $H$ is the same as Bohr's Hamiltonian

\be
H = \frac{\pi^i \pi^i}{2\mu} - \frac{1}{\eta}\ .\label{Bohreq}\ee
(in Bohr's case, of course, the $SO(4)$ generators
are written in terms of coordinates and momenta in real three-dimensional space).

Comparison between Eq.(\ref{lc-commutator}) and Eq.(\ref{llc})  yields the mass squared operator

\be M_{}^2~=~-2\mu \, H\ ,\ee which must then appear in the light-cone Hamiltonian
$$P_{}^- ~ = ~ \frac{p_{}^i p_{}^i +M_{}^2}{2p^+}~=~ \frac{p_{}^i p_{}^i -2\mu \,H}{2p_{}^+}\ .$$
Each mass level has the same degeneracy as that of the hydrogen atom. Two commuting
$SO(3)$'s are generated by the combinations

$$ \frac{1}{2}\left(\,L^k_{} + \frac{T_{}^k}{\sqrt{-2\mu\,H}}\, \right) , \qquad
\frac{1}{2}\left(\,L^k_{} - \frac{T_{}^k}{\sqrt{-2\mu\,H}}\, \right)\ ,  $$
corresponding to  $SO(4)\sim SO(3)\times SO(3)$. Since

\be L_{}^{k}\,T_{}^k~=~\,T_{}^k\,L_{}^{k}~=~0\ ,\ee they have the same Casimir operator

\be
 C^{}_2 ~ = ~ \frac{1}{4}\left(\,L^{k}_{}\,L_{}^k -\frac{T_{}^k\,T_{}^k}{2\mu\,H}\,\right)
~ = ~ j(j+1) \ , \ee
where $j$ is related to the mass operator by

\be
 M_j^2 ~ = ~ \frac{\mu_{}^2}{4C^{}_2 + 1} ~ = ~ \frac{\mu_{}^2}{(\,2j+1\,)_{}^2} \ , \ee
so that $(2j+1)$ is Bohr's principal quantum number. At each mass level, these states assemble in
a representation of $SO(4)$ and generate the spectrum of the infinite component Majorana wave
equation. Since $SO(4) \supset SO(3) \supset SO(2)$, we need to contract at least twice to obtain
a CSR in four dimensions. One contraction parameter is of course the inverse radius of the
compactified direction and the other one is not yet defined. There are two  ways to proceed with
the contractions:
\begin{enumerate}

\item We contract, staying in five dimensions first,  and then use KK reduction and contract to  four
dimensions, i.e.,

$$ SO(4) \xrightarrow[\mbox{ contraction}]{\mbox{1st}} E(3) \xrightarrow[\mbox{reduction}]
{\mbox{KK}} SO(3) \xrightarrow[\mbox{contraction}]{\mbox{2nd}} E(2) \ . $$
\item We apply the KK reduction first and then contract, i.e.,
$$ SO(4) \xrightarrow[\mbox{reduction}]{\mbox{KK}} SO(3) \xrightarrow[\mbox{contraction}]
{\mbox{Double}} E(2) \ . $$
\end{enumerate}
We consider both cases in the following and find the conditions for these procedures to lead to
the CSR in four dimensions.

\subsubsection{Case 1: Contraction to $E(3)$}
In the Majorana theory, the degeneracy at each mass level is generated by the addition of two
angular momenta $j$. The states are eigenstates of their $SO(3)$ diagonal subgroup (the angular
momentum in Bohr's model), so that

\be L^{k}_{}\,L_{}^{k}~=~l(l+1)\ ,\label{ang}\ee where $l=0,1,2,\dots,2j$. This is the analog of
the magnetic quantum number in the $SO(3)\rightarrow E(2)$ contraction process: its range goes to
infinity while its value  remains finite. This leads us to  consider the contraction where all
three $L^k$ that generate $SO(3)$ remain finite as $j\rightarrow\infty$.

In that limit, the mass vanishes and the $T^k$ become commuting translations

\be
 M^{}_j~=~\frac{\mu}{2j+1}~\rightarrow~0\, , \qquad
[\,T_{}^i\,,\,T_{}^k\,]~\rightarrow~0\ , \ee producing $E(3)$, the Euclidean
group in three dimensions. This contraction of the transverse little group
yields a CSR in five dimensions.  To see this explicitly, we multiply the
$SO(4)$ Casimir operator by the contraction parameter $\epsilon$:

\be\nonumber (\epsilon\,L^{k}_{})(\epsilon\,L_{}^k) +\frac{\epsilon^2_{}}{M^{2}_j}\,T_{}^k\,T_{}^k
~ = ~4\,\epsilon^2_{}\, j(j+1) \ . \ee As $\epsilon\rightarrow 0$ and  $j\rightarrow \infty$, for
fixed $\epsilon\,j$ we get

\be
T_{}^k\,T_{}^k~=~\mu^2_{}\ ,\ee
so that the length of the translation vector is fixed.

In five dimensions, the Poincar\'e group has three Casimir operators, the momentum squared (zero
in this case),

\be W^{}_{\mu\,\nu}\,W_{}^{\mu\,\nu} ~~~\text{and}\qquad
W~=~\epsilon^{}_{\mu\,\nu\,\rho\,\sigma\,\lambda}\,P^\rho_{}\,M^{\mu\,\nu}_{}
\,M^{\sigma\,\lambda}_{} \ , \ee where

\be W^{}_{\mu\,\nu}~=~\epsilon^{}_{\mu\,\nu\,\rho\,\sigma\,\lambda}\,
P^\rho_{}\,M^{\sigma\,\lambda}_{}\ . \ee In this case,
\be W^{}_{\mu\,\nu}\,W_{}^{\mu\,\nu}~=~\mu^2_{}\qquad \mbox{and}\qquad W~=~T^{}_k\,L^{}_k~=~0\ .
\ee These two Casimirs characterize a unique CSR in five dimensions. The dimensionality of
space-time does not change, but the contraction parameter $\epsilon$ does not have an obvious
physical meaning.

We can use this contracted theory as a starting point to obtain a four-dimensional theory.
So we apply the KK reduction to the $E(3)$ algebra to get down to four dimensions. The
mass operator and the light-cone vector now takes the following form,

\be M^2_n ~ = ~ \frac{n^2}{R^2} \quad {\mbox{and}} \quad \wt^a ~ = ~ T^a + \frac{n}{R} S^{3a} \ ,
\qquad a=1,2. \ee which can be checked by analyzing the light-cone Hamiltonian and the dynamical
boosts respectively. The generators $S^{12}$ and $\wt^a$ now form $SO(3)$ the massive little group
in four dimensions. The contraction parameter is now $\epsilon' = \frac{1}{R}$. Let $\tj(\tj+1)$
be the eigenvalue of this $SO(3)$ angular momentum algebra. It is quite clear that to get a CSR,
we take the limits,

$$
R\to\infty ,\quad \tj\to\infty \quad \mbox{with} \quad \frac{\tj}{R} \ \
\mbox{fixed} \ .$$
Therefore we obtain,

\be \wt^a \wt^a ~ = ~ \frac{n^2\tj^2}{R^2} ~ \equiv ~ \txi^2 \ , \label{txi-square}\ee
which is the Casimir of the $E(2)$ algebra. The other Casimir is of course the mass
squared which is zero. These two uniquely determines the CSR in four dimensions.

Notice that the first contraction parameter could have been identified with $\frac{1}{j}$ or
$\mu$; both would have yielded a zero mass, but it is only the first case that leads to the CSR;
the other $\mu\rightarrow 0$ generates a normal massless representation of definite helicity.


\subsubsection{Case 2: KK Reduction first} In this case, we apply the KK reduction first.
The massive $SO(4)$ little group in five dimensions now reduces to the massive little group in
four dimensions. The light-cone Hamiltonian becomes

\be P^- ~ = ~ \frac{1}{2p^+}(p_{}^a p_{}^a +M_{jn}^2)\ , \ee
for $a=1,2$, with

\be M^2_{jn}~=~\frac{\mu_{}^2}{(2j+1)_{}^2}+\frac{n_{}^2}{R_{}^2}\ ,\label{m_jn}\ee
using Eq.(\ref{P3}). The generators of the $SO(3)$ little group are

\be L^3 \quad \mbox{and} \quad \wt^a = \left(T^a + \frac{n}{R} S^{3a} \right) \ , \qquad
a=1,2  \ee deduced by looking at $M^{-a}$.

The quadratic Casimir is
\be (L^3)^2  + \frac{1}{M^2_{jn}} \left(T^a + \frac{n}{R} S^{3a} \right) \left(T^a +
\frac{n}{R} S^{3a} \right) ~ = ~ j^\prime (j^\prime +1) \ . \label{4d_angmom}\ee
The mass operator in Eq.(\ref{m_jn}) clearly shows that we need to take the double limit $j,R\to
\infty$ to obtain the massless case. However this is not enough to get a CSR, because from
Eq.(\ref{4d_angmom}), it is also clear that the ratio $\frac{j^\prime}{R}$ must also remain finite
and nonzero. To see it explicitly, we divide Eq.(\ref{4d_angmom}) by $R^2$ and rearrange to get
\be \wt^a \wt^a ~ = ~  \left( \frac{\mu_{}^2}{(2j+1)_{}^2/R^2} + n_{}^2 \right) \left(
\frac{j^\prime}{R} \right)^2 - M^2_{jn} \, m^2 , \ee
where we have used $L^3=m$. The double limit certainly does not give any nonzero value, but the
following triple limit

\be j, j^\prime, R \to\infty \ , \quad \mbox{such that} \quad \frac{j}{R} , \ \frac{j^\prime}{R}
\quad \mbox{remain finite (non-zero)} \ , \ee does give a finite value. Therefore, in this triple
limit, we find,

\bea (\widehat T^a )^2 & = & \left( \frac{\mu^2}{(2j/R)^2} + n^2
\right)\left(\frac{j^\prime}{R}\right)^2 ~ \equiv ~ {\Xi^\prime}^2 \ ,
\label{pxi-square}\eea
which is the Casimir in four dimensions. This and the mass squared (which is zero) uniquely labels
the representation. The difference between Case 1 and Case 2 is the length of the light-cone
vector. This features is what we expected, because the generator $S^{12}$ was not affected by any
of the reduction and/or contraction processes. The contraction only affected the quadratic Casimir
eigenvalues which sets the range of quadratic Casimir of its maximal subgroup. Therefore, the
representations of the contracted algebra of these two cases should differ by the length of the
light-cone vectors and that's what we have found. These representations are also consistent with
In\"on\"u and Wigner's result with different length of the Euclidean light-cone vectors.
%
%
\subsection{Representation of the Wavefunction}
Since we have the form of the $T^i$s in terms of the internal variables, we can explicitly find
the representation of the wave function for the light-cone vectors $\wt^a$. First note that in
$\wt^a$ the contraction parameter $j$ is not explicitly present. Therefore if we apply the
contraction it will only include the $R\to \infty$ limit, not the $j\to\infty$. As a result we
will not obtain a commuting vector which is required to obtain the CSR. Since the contraction
parameter $j$ appears through the Hamiltonian when it acts on the wavefunction, we rewrite $T^i$
in terms of the Hamiltonian as in the following,

\bea T^i & = & \epsilon_{ijk} \pi^j L^k - \frac{\mu \eta^i}{\eta} - i \pi^i \ , \nonum \\
& = & \eta^i \left( 2 \mu H + \frac{\mu}{\eta} \right) - ( \eta \cdot \pi - i )\pi^i \ ,
\eea
where we used $H$ as given by Eq.(\ref{Bohreq}). Therefore the light-cone
vector, $\wt^a$, can be expressed as

\be \wt^a ~ = ~ \left( 2 \mu \eta^a  H + \frac{\mu \eta^a }{\eta} - \frac{n}{R} \eta^a \pi^3  -
\mu \left[ \, \eta^a \, , \, H \, \right] \right) + ( y + 2i - \eta \cdot \pi) \pi^a \
,\label{wt-def} \ee where \be y~=~\frac{n\,\eta^3}{R}\ .\ee The first term is dependent on the
contraction parameters whereas the second term is independent and also both term are separately
hermitian. Since $H \sim {\cal O}(1/j^2)$ and $1/\eta \sim {\cal O}(1/R)$, the first term vanishes
as $j,R \to \infty $.

 It should be noted that if we evaluate the commutator in the first term in
Eq.(\ref{wt-def}) and then consider the contraction, the resulting light-cone vector
becomes non-hermitian and the whole analysis becomes physically meaningless. It is not
well understood why this happens, but to maintain the hermiticity, we have to consider
the contraction without computing the commutator. Using
$$\eta \cdot \pi = \eta^a \pi^a - i y \frac{\partial}{\partial y} \ ,$$
and dropping the first term, the light-cone vector becomes,
\be \wt^a ~ = ~ \left(y + 2i + iy \frac{\partial}{\partial y} - \eta^b \pi^b \right) \pi^a \ . \ee
It can be easily checked that the above form satisfies the $E(2)$ algebra,
$$ \big[ \, \wt^a \ , \ \wt^b \, \big] ~ = ~ 0 \quad \mbox{and} \quad \big[ \, S^{ab} \ , \
\wt^c \, \big] ~ = ~ i\left( \eta^{ac}\, \wt^b - \eta^{bc}\, \wt^a \right) \ , $$
which is required to obtain any CSR in four dimensions. We now find the representation of
the wavefunction for this $E(2)$ algebra.

The square of the light-cone vector is
\bean (\wt^a)^2 & = & \left[ (\eta^b \pi^b)^2 - i\eta^b \pi^b \left( -2iy +5 + 2y \frac{d}{d y}
\right) \right. \nonum \\ & & \quad \left. - \left( y^2 \frac{d^2}{dy^2} - (2iy^2 -6y)\frac{d}{dy}
-( y^2 + 6iy -6)\right) \right] (\pi^2) \ . \eean
On a wavefunction of the form,
$$ \Phi(\eta^i) ~ = ~ e^{iy} \phi(\eta^a , y) \ , $$
this gives the following form of the
differential equation,
\be \left[ (\eta^b \pi^b)^2 - i\eta^b \pi^b \left( 5 + 2y \frac{d}{d y} \right) - \left( y^2
\frac{d^2}{dy^2} + 6y\frac{d}{dy} + 6)\right) \right] (\pi^2)\, = \, \Xi^2 \ ,
\label{wt-square}\ee
where $\Xi=\tilde\Xi$ or $\Xi'$ corresponding to cases 1 and 2, respectively. To solve the above
equation, let
\be (\pi^a)^2 ~ = ~ \Pi^2 (y) \, \ee when it acts on $\phi(\eta^a , y)$, that is the
representation of the $\eta^a$ dependent part of the wave function is the Bessel function, namely,

$$ \phi(\eta^a , y) ~ = ~ J_m(\Pi(y) \rho) e^{im\theta} F(y)\ , $$
where $F(y)$ is a real function and we used the polar coordinates, $\eta^1 = \rho \cos
\theta$ and $ \eta^2 = \rho \sin \theta$. In polar coordinates, we use,
\bean - i \eta^a \pi^a & = & - \rho \frac{d}{d \rho} \ , \\ (\eta^a\pi^a)^2 & = & -
\left( \rho^2 \frac{d^2}{d \rho^2} + \rho \frac{d}{d \rho} \right) \ , \eean
and we rewrite Eq.(\ref{wt-square}) as

\bea \left[ \rho^2 \frac{d^2}{d\rho^2} + \left( 6 + 2y\frac{d}{dy} \right) \rho
\frac{d}{d\rho} + \left( y^2 \frac{d^2}{dy^2} + 6y \frac{d}{dy} + 6 \right) \right]
\left( \Pi^2 J_m  F  \right) \, = \, -\Xi^2 J_m  F \ , \label{diff-eqn}\eea
Making use of the following identities,

\bea \frac{d}{d\rho} J_m(\Pi(y) \rho) & = & \frac{\Pi}{2} \left( J_{m-1} - J_{m+1}
\right) \ , \\ \frac{d}{dy} J_m(\Pi(y) \rho)& = & \frac{\rho}{2} \frac{d\Pi}{dy} \left(
J_{m-1} - J_{m+1} \right) \ , \eea
and after straightforward algebra, the left-hand side of Eq.(\ref{diff-eqn}) can be
expressed as a linear combination of Bessel function of different orders,

\bean & & \left(\frac{\rho^2 \Pi^4 F}{4} + \frac{\rho^2 y \Pi^3 \Pi^\prime F}{2}
+\frac{\rho^2 y^2 \Pi^2 {\Pi^\prime}^2 F}{4} \right)\Big( J_{m-2} + J_{m+2} \Big)
\\ & + & \Bigg( 3 \rho \Pi^3 F + 6 \rho y \Pi^2 \Pi^\prime F + \rho y \Pi^3 F^\prime +
2 \rho y^2 \Pi {\Pi^\prime}^2 F \\ & & \quad + \rho y^2 \Pi^2 \Pi^\prime F^\prime
+ \frac{\rho y^2 \Pi^2 \Pi^{\prime \prime} F}{2} \Bigg)\Big( J_{m-1} - J_{m+1} \Big) \\
& +& \left( - \frac{\rho^2 \Pi^4 F}{2} - \rho^2 y \Pi^3 \Pi^\prime F + 2y^2
{\Pi^\prime}^2 F + 2y^2 \Pi \Pi^{\prime \prime} F + 4y^2 \Pi \Pi^\prime F^\prime \right.
\\ & & \quad \left. + y^2 \Pi^2 F^{\prime \prime} - \frac{\rho^2 y^2 \Pi^2 {\Pi^\prime}^2 F}{2} +
12 y \Pi \Pi^\prime F + 6y \Pi^2 F^\prime + 6 \Pi^2 F \right) J_m  \ , \eean
where the `prime' denotes derivative with respect to the argument. By matching the coefficients of
Bessel functions of different order with the r.h.s. of Eq.(\ref{diff-eqn}), we get three
constraints. Equating the coefficient of $J_{m\pm 2}$ to zero gives the first constraint,

$$ \frac{\rho^2 \Pi^2}{4} \left( \Pi^2 + 2y \Pi \frac{d\Pi}{dy} + y^2 \left( \frac{d\Pi}{dy}
\right)^2 \right) F(y) ~ = ~ 0 \ , $$
which is satisfied if

\be \Pi^2 + 2y \Pi \frac{d\Pi}{dy} + y^2 \left( \frac{d\Pi}{dy} \right)^2 ~ = ~ 0 \ .
\label{pi-eqn} \ee The solution to this equation is given by
\be \Pi(y) ~ = ~ \frac{\Pi_0}{y} \ , \label{piy} \ee where $\Pi_0$ is a constant. The constraint
obtained by equating the coefficients of $J_{m\pm1}$ provides no new result. Finally the remaining
constraint is from the coefficients of $J_m$ which gives a differential equation for $F(y)$,

\be \frac{d^2 F}{dy^2} + \frac{2}{y} \frac{dF}{dy} + \frac{\Xi^2}{\Pi_0^2} \, F ~ = ~ 0 \ , \ee
where we used $\Pi(y) = \Pi_0/y$ and its derivatives. The solution is given by,

\be F(y) ~ = ~ \frac{1}{y} \left( A \sin\left(\frac{\Xi y}{\Pi_0} \right) + B \cos \left(
\frac{\Xi y}{\Pi_0} \right) \right)\ , \ee where $A$ and $B$ are constants, and the regularity
condition at $y=0$ implies $B=0$. Therefore the complete wave function can be written as,
\be \Phi(\rho,\theta,y) ~ = ~ \sum_m N_m \left(\frac{e^{iy}}{y} \sin\left(\frac{\Xi \, y}{\Pi_0}
\right) \right) e^{im\theta} J_m\left(\frac{\Pi_0 \rho}{y}\right) \ , \label{csr-sol}\ee where
$N_m$ is the overall normalization constant.

In their paper\cite{iw}, IW found the following wavefunction for the $E(2)$ algebra,
\be \Phi_{IW}(\rho, \theta) ~ = ~ \sum_{m} c_m \, e^{im\theta}\, J_m(\Xi \rho)
\ , \label{wigner-sol}\ee
where $\Xi^2$, the square of the Pauli-Lubanski vector, is the second Casimir of the Poincar\'e
group, and labels the CSR in four-dimensions. The Euclidean vector is linear in $\pi^a$ and as a
result its length appeared as a scale factor in the Bessel function. In four dimensions there are
only two types of CSRs, fermionic and bosonic types corresponding to half-odd and integer values
of $m$. The amplitude is also constant.

Although similar in form, there are differences between Eq.(\ref{csr-sol}) and the IW form in
Eq.(\ref{wigner-sol}). In our case, the internal momentum (Eq.(\ref{piy})) and amplitude are not
constant, but functions of $y$ which is the ratio of the internal and external coordinates. It is
due to the fact that the Euclidean light-cone vector $\wt^a$ is not linear in $\pi^a$ (in addition
to the $y$ dependence). Moreover, unlike the IW case, the length of the light-cone vector does not
appear as a scale factor in the Bessel function, even though it is the CSR. To find how our CSR is
related to IW's CSR, we must find a relation between $\Pi$ and $\Xi$. Let's assume
$\Pi=\Xi=\Pi_0/y$. The only solution we get, following Eq.(\ref{pi-eqn}), is $\Xi=0$ which
corresponds to the regular massless representation, not the CSR. Therefore $\Pi\ne \Xi$ is the
only possibility and these are quite new. On the other hand, we may assume that, instead of $\Pi$,
$\Pi_0=\Xi$. Substituting this into Eq.(\ref{csr-sol}) and setting $y=1$, the solution becomes
exactly the same as that of Wigner's CSR apart from the overall phase factor which has no physical
effect (the overall constant factor can be absorbed into the normalization factor). For any other
values of $y$, we have a different kind of CSR because of the non-equality between the scale
factor and length of the Euclidean vector. There is no physical reason for $\Pi_0$ and $\Xi$ to be
equal and $y=1$, but this is the only condition to obtain IW's result.

Finally, the raising and lowering operators are defined as,
$$ \wt^\pm ~ = ~ \wt^1 \pm  i\wt^2 \ . $$ In polar coordinates, these become
\bea \wt^+ & = & e^{i\theta} \left( 2 + y\frac{d}{dy} + \rho \frac{d}{d\rho}\right) \left(
\frac{d}{d\rho} + \frac{i}{\rho}\frac{d}{d\theta} \right) \ , \\ \wt^- & = & e^{-i\theta} \left( 2
+ y\frac{d}{dy} + \rho \frac{d}{d\rho} \right) \left( \frac{d}{d\rho} -
\frac{i}{\rho}\frac{d}{d\theta} \right) \ , \eea which acts on $\phi(\rho, \theta, y)$. It is
quite obvious that the states $\wt^+|>$ and $\wt^-|>$ have helicities $(m+1)$ and $(m-1)$
respectively. The remaining generator $T^3/R$ becomes
$$ \frac{T^3}{R} ~ = ~ (\pi^a)^2 \ , $$
under contraction. The complete Poincar\'e wavefunction is,
$$ \Phi(x^+, x^a; \rho,\theta, y) ~ = ~ \sum_m N_m \frac{e^{iy}}{y} \sin\left( \frac{\Xi
y}{\Pi_0}\right) \, e^{-i(x^- p^+ - x^a p^a)} \, e^{im\theta} J_m \left( \frac{\Pi_0
\rho}{y} \right) \ . $$

The physical meaning of the new parameter $y$ is not clear to us, but it links the external to the
internal coordinates of the Majorana theory. This type of limits links the internal structure  to
that of space-time. If we view, as we have, the Majorana theory as a warm-up for string theory, it
may correspond to compactifying a string with its vibrational modes stuck along an extra
dimension. This may help define the infinite slope limit of string theory.

%
\section{Conclusion}
In this paper, we showed that if we dimensionally reduce from five to four dimensions, the
compactified algebra is the regular representation of the Poincar\'e group in four dimensions with
mass $M_n$ for all modes. For the infinite radius limit, we find that the KK mass spectra
collapses to the lowest mass state and gives the regular representation of the Poincar\'e group in
four dimensions. Only for non-zero modes and infinite radius limit, we obtain CSR if the
$\frac{j}{R}$ remains constant as $R\to \infty$, otherwise we will get the regular representation.

We constructed explicitly the representation of the Euclidean group in a system with periodic
boundary condition for internal coordinates and found the condition to obtain non-zero Euclidean
vectors using IW contraction method. We found that for nonzero and finite Euclidean vector, the
eigenvalue of the uncontracted group must be proportional to the radius of the compactified
direction. This condition is similar to that of the CSR where the spin is proportional to the
radius, apart from the mode number. Therefore finding the representations of any Euclidean group
provides the representations of the inhomogeneous light-cone little group which is the CSR.

We have applied the contraction technique to both regular massless and massive representations in
five dimensions. In the regular massless case, the contraction yields the $E(2)$ little group
under the double limits $j,R\to\infty$, keeping $j/R$ finite and non-zero.

The massive case  is not as straightforward as the regular massless case. We have considered
Majorana theory as a model in the five dimensional massive case. We have found that the double
limits are not enough to get a CSR, even though the mass operator vanishes. We have to consider
triple singular limits $j,j^\prime,R\to\infty$ such that both $j/R$ and $j^\prime/R$ remain fixed
and non-zero. There are two ways to consider the contraction limits, in the sequence of
contraction, KK reduction and contraction, or KK reduction and double contraction. Both of these
yield the CSR in four dimensions. The difference is in the length of the light-cone vector. The
representation wavefunction of the contracted algebra is the Bessel function, however as in
Wigner's solution, the length of the light cone vector does not appear as a scale factor in the
Bessel factor. In our case both the amplitude and the scale factor are a function of the parameter
$y$ which is a ratio of the internal and external coordinates. We have found that the contracted
algebra is identical to that of Wigner's CSR only if $\Pi_0$ which is the magnitude of the
internal momenta at $y=1$, equals the length of the light-cone vector (apart from an overall phase
factor).

Thus, the CSR can arise in the Majorana theory, but with an extra variable. Its physical
interpretation may provide clues when we apply our techniques to string theory in the hope of
identifying the CSR with its infinite slope limit.

Our technique may also prove useful when the starting point is curved space in higher dimensions,
particularly Anti-de-Sitter. There the starting group is the Conformal, not the Poincar\'e  Group.
This possibility, though interesting in its own right, will not be discussed here.

\section{Acknowledgements}
We wish to thank Professor L. Brink for many enlightening discussions in the course of this work.
This Research is supported in part by the US Department of Energy under grant DE-FG02-97ER41029.
\newpage
\begin{appendix}
\noindent{\Large{\bf Appendix: In\" on\" u-Wigner \`a la Kaluza Klein}} \vskip .2cm \noindent In
this appendix we reproduce their results by a different method. We  consider the full $SO(3)$
algebra which acts on three-dimensional vector space with periodic boundary condition along one
direction The contraction parameter in  is the inverse radius of that direction. After the
contraction, the three-dimensional wavefunction reduces to  Bessel functions in the radial
coordinate. In the following we retrace IW's steps, stressing the geometrical  picture of the
contraction procedure.

The IW contraction of $SO(3)$  to $E(2)$ amounts to the study of  a dynamical system with $SO(3)$
symmetry restrained to a space whose  boundary condition breaks that symmetry. We switch to
cylindrical coordinates, and seek solutions which are periodic in $z$,
\bea \label{z-periodic}z & \rightarrow & z+ 2\pi R \ , \\ \Phi(\rho ,\theta , z) & = &
\sum_n \Phi_n(\rho ,\theta) e^{inz/R} \ , \label{wf-periodic}\eea
where $n$ is the mode number. In the following we will use $1/R$ as the contraction
parameter. The  angular momentum operators are given by,
\bea S^{12} & = & -i \frac{\partial}{\partial \theta} \  , \label{s12} \\ S^{23} & = & - i \left(
-z \sin\theta \frac{\partial}{\partial \rho} + \rho \sin\theta \frac{\partial}{\partial z} -
\frac{z \cos\theta}{\rho} \frac{\partial}{\partial \theta} \right) \ , \label{s23} \\ S^{31} & = &
-i \left(z \cos\theta \frac{\partial}{\partial \rho} - \rho \cos\theta \frac{\partial}{\partial z}
- \frac{z \sin\theta}{\rho} \frac{\partial}{\partial \theta} \right) \ . \label{s31} \eea
Acting on the wave function, the factor $\frac{\partial}{\partial z}$ can be replaced by
$\frac{in}{R}$. When $R$ is very large (ultimately we will take it to $\infty$), we  drop
all terms proportional to $ \frac{\partial}{\partial z}$, yielding
\bea S^{31} & \to & - iz \left( \cos \theta \frac{\partial}{\partial \rho} -
\frac{\sin\theta}{\rho} \frac{\partial}{\partial \theta}\right) \ , \\ S^{32} & \to & iz \left(
-\sin\theta \frac{\partial}{\partial \rho} -\frac{\cos\theta} {\rho} \frac{\partial}{\partial
\theta} \right) \ . \eea
However $z$ is no longer well defined on the wave function. The Casimir operator  now becomes
\bea  (S^{12})^2+(S^{31})^2 + (S^{32})^2 &=& j(j+1)\ ,\\ &\to&   -z^2
\left(\frac{\partial^2}{\partial \rho^2} + \frac{1}{\rho} \frac{\partial}{\partial \rho} +
\left(\frac{1}{\rho^2} + \frac{1}{z^2} \right) \frac{\partial^2}{\partial \theta^2} \right) \ .
\label{e2operator}\eea
The wave function on which this differential operator acts is of the form,
$$ \Phi (\rho ,\theta)
~ = ~ {\cal R}(\rho) \Theta(\theta) \ , $$
with eigenvalue $s^2$. Since there is no mode dependence, we have dropped the subscript $n$ from
the wavefunction. For the above form of the wavefunction, the angular part decouples from the
radial part and can be written as
\be  \frac{1}{\Theta} \frac{d^2\Theta}{d \theta^2} ~ \equiv ~ - m^2 \ , \ee where $m$, the
azimuthal quantum number, is an integer or half-odd integer depending on whether the wave function
is single-valued or double-valued. Let's also define the dimensionless variable
$$ \trho ~ = ~ \frac{\sqrt{s^2 -m^2}}{z}\, \rho~\equiv~{q}(s,m) ~\rho  \ . $$
Substituting the angular part, the eigenvalue equation for the differential operator in
Eq.(\ref{e2operator}) becomes  Bessel's differential equation
\be \trho^2 \frac{d^2 {\cal R}}{d \trho^2} + \trho \frac{d {\cal R}}{d \trho} + \left(\trho^2 -
m^2 \right) {\cal R} ~ = ~ 0 \ ,   \ee
whose solutions are well-known. The regularity condition at the origin implies that the
physically relevant solution is the Bessel function $J_m(\trho)$. Therefore the wavefunctions for
$E(2)$ group are linear combinations of  the form
\be \label{e2wavefunction}\Phi(\rho ,\theta) ~ = ~  J_m({q}\,\rho ) e^{im\theta} \ . \ee We let

\be T^a_{}~\equiv~\lim_{R\to\infty} \frac{S^{3a}}{R}\ ,\ee that is

$$T^1_{}~=~ \left( \cos
\theta \frac{\partial}{\partial \rho} - \frac{\sin\theta}{\rho} \frac{\partial}{\partial
\theta}\right) \ ,\qquad T^2_{}~=~\left( -\sin\theta \frac{\partial}{\partial \rho}
-\frac{\cos\theta} {\rho} \frac{\partial}{\partial \theta} \right) \ .$$

It can easily be checked that  the $T^{a}$, $a=1,2$,  commute and that they transform as a
2-vector under $S^{12}$, forming a representation of the  $E(2)$ algebra. The length of the
Euclidean vector is now

\be \xi^2_{}~\equiv~T^a~T^a_{} ~ = ~  -  \left(\frac{\partial^2}{\partial \rho^2} + \frac{1}{\rho}
\frac{\partial}{\partial \rho} + \frac{1}{\rho^2} \frac{\partial^2}{\partial \theta^2} \right) \
.\label{e2vector}\ee
The  wavefunction for the above differential equation is  the same as in Eq.(\ref{e2wavefunction})
with scale factor $q$ defined as
\be q(\xi) ~ = ~ \frac{\xi}{z / R} \ , \label{E2xi}\ . \ee
Since in the infinite radius limit these two scale factors are the same, the eigenvalues, $\xi$
and $s$, must be related by \be \xi ~ = ~ 2 \pi q ~ = ~ \frac{s}{R} \ . \ee
This solution is of course the same as that of In\"on\"u and Wigner, because $s$ and $j$ are same
when both are large. The generalization of the above construction to any dimensions is also quite
straightforward.
\end{appendix}

\end{document}